\documentclass[a4paper,12pt]{article}
\newcommand{\beq}{\begin{equation}}
\newcommand{\eeq}{\end{equation}}
\newcommand{\eeqr}{\end{eqnarray}}
\newcommand{\beqr}{\begin{eqnarray}}
\begin{document}
\date{}
\begin{center}
 \Large\bf Black Hole Complementary Principle and The Noncommutative Membrane\\

\end{center}
\vspace*{1.2cm} \centerline{\large Zen Wei$^{\dagger\ddagger}$}
\begin{center}
\vspace*{0.7cm}
\medskip
$^{\dagger}$\emph{Institute of Theoretical Physics, Chinese Academy of Science  \\
\it  P. O. Box 2735 \it Beijing 100080 \it P.R. China} \\
\medskip
$^{\ddagger}$\emph{Graduate School of the Chinese Academy of Sciences\\Beijing 100080, P.R. China}\\
\vspace*{0.7cm}
{\tt vision@itp.ac.cn}
\end{center}

\vspace*{1.8cm}

\centerline{\bf Abstract}In the spirit of Black Hole Complementary
Principle, we have found the noncommutative membrane of
Scharzchild Black Holes. In this paper we extend our results to
Kerr Black Hole and see the same story. Also we make a conjecture
that spacetimes is noncommutative on the stretched membrane of the
more general Kerr-Newman Black Hole.

\vspace*{0.5cm} \vfill \noindent {\bf November 2005}
\thispagestyle{empty} \setcounter{page}{0}
\newpage
\section{Introduction}
To solve the black hole information loss problem \cite{1}, L.
Susskind found the Black Hole Complementary Principle \cite{2},
which say that there's no such super-observer who can see both the
inside and outside of black holes, and what the inside and outside
observers see are complementary. From this principle what the
outside observer see is self-consistent, and there is no black
hole information loss for him because of the black hole
information red-shift on the horizon and we can take the black
hole as a physical object.

For the outside observer who is near the horizon, the black hole
behaves like a hot stretched membrane.  This stretched membrane is
much different from its original meaning which is just a classical
fictitious construct \cite{3}. In fact for this observer, all the
quantum information of the black hole is stored on the hot
membrane and we can get some insight from the membrane \cite{3-1}.

In our previous paper \cite{4}we have found the unusual dispersion
relation $E=\frac{m^2}{k}$ on the stretched membrane, which
indicates the noncommutative spacetimes on the membrane for the
Scharzchild Black Hole. In the spirit of Black Hole Complementary
Principle we extend our results to Kerr Black Hole and see the
same story. Also we make a conjecture that spacetimes is
noncommutative on the stretched membrane of the more general
Kerr-Newman Black Hole.

In this paper we will first give a review of our results for the
Scharzchild Black Hole, and the extended results for Kerr black
holes will be given next. Last we will show some arguments to make
our conjecture.

\section{Noncommutative membrane of Scharzchild Black Holes}

In this section we will give a brief review of our previous
results for Schwarzschild black holes. The 4 dimensional black
hole with the Schwarzschild metric takes the form:
 \beq ds^2=-(1-{r_0\over r})dt^2+(1-{r_0\over
r})^{-1}dr^2+r^2d\Omega^2_2 \eeq For the infinite observer the
Hawking temperature and the Bekenstein-Hawking entropy are \beq
T_H={1\over 4\pi r_0},\ S=\frac{A}{4l_p^2},\ \ \ \ l_p^2=G）\eeq
where $A$ is the area of the horizon. For the observer near the
horizon the black hole looks like a hot stretched $2+1$
dimensional membrane and the Hawking temperature will be
blue-shifted: \beq T=g^{-1/2}_{00}(r_s)T_H \eeq If we attribute
the black hole mass completely to the stretched horizon, since the
mass is also blue-shifted, the mass density on the stretched
horizon is \beq \rho=Mg^{-1/2}_{00}(r_s)/A \eeq The density of
entropy is simply \beq \sigma=S/A=\frac{1}{4l_P^2} \eeq For the
Schwarzschild black hole the Smarr' mass formulas \cite{5} is \beq
M=2T_HS \eeq So from above relations we can easily get \beq
\label{r} \rho=2T\sigma \eeq As was stated in our previous paper
the above formulas means that the energy per degree of freedom is
$2T$. It's a  strange result because the energy per degree is
$T/2$ for the one-dimensional relativistic gas, and for a $d$
dimensional relativistic gas, we have $\rho/\sigma =(d/(d+1))T$.
Thus, we can never hope to get the result $\rho/\sigma=2T$ from
any known gas.

Let us assume that on the membrane there is a perfect fluid. We
shall for now imagine that its temperature, energy density and
entropy density can be changed, and then we can apply the first
law of thermodynamics to the membrane to gain some insight into
this fluid: \beq d(\rho A)=Td(\sigma A)-pdA \eeq So we get \beq
A(d\rho -T\sigma)+(\rho -T\sigma +P)dA =0 \eeq Because of the
symmetry of the Schwarzschild metric, the thermodynamic quantities
measured by the observer near the horizon should be independent of
$A$. Therefore we obtain \beq \label{rr} \rho=T\sigma -p,\quad
d\rho = Td\sigma  \eeq From the first relation and (\ref{r}) we
get \beq p=-T\sigma =-\rho /2 \eeq Thus the pressure is negative
which may be responsible for the stability of the membrane under
self-gravity. From (\ref{r}) and the second relation we get \beq
\label{rrr} \rho =2cT^{-1}\ \ \ ,\sigma =cT^{-2} \eeq Obviously
the above relation (\ref{rrr}) is unusual.

 Next we try to use statistical mechanics to gain some insight into the microscopic
nature of the membrane fluid. All thermodynamic quantities can be
obtained from the free energy density $f$  \beq
\rho=\partial_\beta f,\quad \sigma =\beta \rho -f \eeq  To obtain
(\ref{rrr}) we should require that $f$ scales with $T$ as $f \sim
\beta ^{2}$. For the noninteracting relativistic gas we have \beq
\label{w} f=\int_0^\infty F(\beta k)kdk =\int_0^\infty \mp\ln(1\pm
\exp(-\beta k))kdk \eeq where $F$ as a function of $\beta k$
depends on the nature of the constituents, bosons or fermions and
the measure $kdk$ in the momentum space is standard for a 2
dimensional surface. In order to have $f\sim T^{-2}$, one can
either replace the measure $kdk$ by $k^{-3}dk$, or to change the
dispersion relation.  We have no reason to change the measure (a
consequence of quantum mechanics). Thus, we need to change the
dispersion relation $E=E(k)$ in $F(\beta E(k))$. It is easy to see
that if the dispersion relation satisfies  \beq \label{x}
E=\frac{m^2}{k} \eeq, we get the desired result $f\sim T^{-2}$.
 Thus in conclusion, the following are forced upon us \beq
f=\int_0^\infty F(\beta m^2/k)kdk =am^4\beta^2 \eeq \\ with \beq
\label{x} a=\int_0^\infty F(k^{-1})kdk \eeq

We should remember that that the usual quantum mechanics is still
valid on membrane, so momentum is still conjugate to space, and
time is conjugate to energy, thus $ E\sim 1/\Delta T$ and $k \sim
1/\Delta X $. Thus the unusual dispersion relation
$E=\frac{m^2}{k} $ naturally implies \beq  \Delta T \Delta X \sim
m^{-2} \eeq  Therefore we have found that the spacetimes on the
stretched membrane is noncommutative \cite{6}. In our previous
paper another interpretation was given that the constituents of
the membrane fluid is microscopic black holes. But as we will show
in the next section, this picture may be wrong because all the
thermodynamic quantities on the stretched membrane depend on the
local observer and have just local meaning, and so the global
interpretation should be meaningless that the constituents of the
membrane fluid is microscopic black holes.

\section{Noncommutative membrane of
Kerr Black Holes} Intuitively we can guess that for the local
property of the stretched membrane there should be no difference
between the Scharzchild and the Kerr black holes if the observer
comoves with the horizon. In the following we can see this point.

The kerr black hole metric in Boyer-Lindquist coordinate$(
x^{0}=t,x^{i}=r,\theta^{\dag}, \phi^{\dag} )$, is given by \beq
ds^{2}=-( \rho^{2} \Delta/ \Sigma^{2} )dt^{2} +g_{jk}( dx^{j}+
\beta^{k}dt )( dx^{k}+\beta^{k}dt ), \eeq where \beq
g_{jk}dx^{j}dx^{k}=( \rho^{2}/\Delta
)dr^{2}=\rho^{2}d\theta^{\dag2}+(\Sigma
sin\theta^{\dag}/\rho)^{2}d\phi^{\dag2} \eeq \beq \Delta \equiv
r^{2}+a^{2}-2Mr,~ ~ \rho^{2}\equiv
r^{2}+a^{2}cos^{2}\theta^{\dag},
~~\Sigma^{2}\equiv(r^2+a^2)^{2}-a^{2}\Delta sin^{2}\theta^{\dag}
\eeq and \beq \beta^{\phi^{\dag}}=-2Mar/\Sigma^2,~ ~ \beta
^r=\beta^{\theta^{\dag}}=0 \eeq The event horizion is at \beq
r_{H}=r_{+}~,~~ r_{\pm}\equiv M\pm (M^2-a^2)^{1/2}, \eeq where
$r_{\pm}$ are roots of $\Delta=0$

Since the strethed horizon is very close to $r_{H}$,we can arrive
at a more convenient set of coodinate \beq \alpha \equiv \rho
\Delta^{1/2}/\Sigma =\left[
\frac{1-a^{2}sin^{2}\theta^{\dag}/2Mr_{H}}{Mr_{H}/(r_{H}-M)}
\right]^{1/2}(r-r_{H})^{1/2}+O((r-r_{H})^{3/2}), \eeq which
vanishes at the event horizon and increases outward. Next we
introduce the  new angular coordinates
$\theta^{\prime}$,$\phi^{\prime}$by \beq \label{new}
\theta^{\prime}=\theta^{\dag}-\frac{\rho_{H,\theta^{\dag}}^{2}}{\kappa^{2}\rho_{H}^{4}}\alpha^{2}
,~ ~ \phi^{\prime}=\phi^{\dag}- \Omega_{H}t \eeq where
$\Omega_{H}$ is the angular velocity of the horizon with
$\Omega_{H}=a/2Mr_{H}$ and
$\rho_{H}^{2}=r_{H}^{2}=a^{2}cos^{2}\theta^{\dag}$ is the value of
$\rho^2$ at the horizon. In terms of these coordinates the
spacetime metric takes the form \beq \label{m1}
ds^{2}=-\alpha^{2}dt^{2}+\kappa^{-2}d\alpha^{2}+\rho_{H}^{2}d\theta^{2}+\tilde{\omega}_{H}^{2}(
d\phi^{\prime}+\beta^{\theta^{\prime}}dt )^{2} \eeq \beq \kappa
\equiv{\frac{r_{H}-M}{2Mr{H}}},~ ~ \tilde{\omega}_{H}^{2}  \equiv
{\frac{(2Mr{H})^{2}}{\rho_{H}^{2}}sin^{2}\theta^{\prime}},\eeq
\beq \beta^{\theta^{\prime}}\equiv {\frac{\alpha^{2}a}{\kappa
(2Mr{H})^{2}\rho_{H}^{2}} \left[
\rho_{H}^{2}r_H+M(r_{H}^{2}-a^{2}cos^{2}\theta^{\prime} ) \right]
} \eeq One advantage of the new angular variables is that they
comove with the horizon. Because we focus our attention on the
strethed horizon ($\alpha\ll1$) which is sufficiently close to the
event horizon then in Eq.(\ref{m1}) the $g_{t\phi^{\prime}}$ term
, which is $O(a\alpha^{2}sin^{2}\theta^{\prime})$, may be ignored
and the kerr geometry may be approximated as \beq
 \label{1111}
ds^{2}=-\alpha^{2}dt^{2}+\kappa^{-2}d\alpha^{2}+\rho_{H}^{2}d\theta^{2}+\tilde{\omega}_{H}^{2}d\phi^{\prime2}
\eeq In the membrane paradigm the surface at
$\alpha=\alpha_{H}\ll1$ is called the ``stretched horzion" which
is timelike. The physical variables such as the temperature $T_s$,
mass density $\rho_{o}$ on the two dimensional membrane are
blue-shifted, which are given by \beq
\rho_{o}=\frac{1}{\alpha_{H}}\rho=
\frac{1}{\alpha_{H}}\frac{M}{A},\
T_{s}=\frac{1}{\alpha_{H}}T_{H}=\frac{1}{\alpha_{H}}\frac{\kappa}{2\pi}
\eeq Also on the membrane, the density of entropy is \beq
\sigma=\frac{S}{A_{s}},\ S=\frac{A}{4},\ A_{s}\simeq{A} \eeq

Obviously $T_{s}$ and $\rho_{o}$ have only local meaning which
depend on $\theta^{\prime}$ and $r$. Also we should realize that
in the above definition of the mass density of the stretched
membrane, the observer is obscure and we just give a blue-shift
factor before the physical quantities measured by the infinite
observer. From the new coordinates (\ref{new}) and the metric
(\ref{1111}) we can learn that the observer co-rotates with the
horizon. So the above definition of the mass density is just a
formal one and its physical meaning is obscure. To get the
physical mass density $\rho_{s}$ measured by the observer who is
near the horizon and comoves with it, we should remove the
contribution of the angular momentum to the formal one $\rho_{o}$.
Also we should notice that the local temperature $T_{s}$ is
physical because it is a macroscopic thermodynamic quantity and
should be independent of the angular momentum.

From Smarr's mass formula for kerr black hole : \beq
M=\frac{A\kappa}{4\pi}+2\Omega_{H}J \eeq where \beq
\kappa=\frac{r_{+}-r_{-}}{4Mr_{+}},\ A=4\pi(r_{H}^{2}+a^{2}),\
\Omega_{H}=\frac{a}{2Mr_{+}} \eeq we can obtain \beq
\rho_{o}=2T_{s}(\sigma+\frac{2\pi\Omega_{H}J}{\kappa A}) \eeq As
was argued in the above, we should  remove the contribution of the
angular momentum to the formal one $\rho_{o}$. So the physical
mass density $\rho_{s}$ measured by the observer comoving with the
horizon should be \beq \label{r1} \rho_{s} =\rho_{o} - 2T_{s}
\frac{2\pi\Omega_{H}J}{\kappa A} = 2T_{s} \sigma \eeq As we see,
for the local physical quantities relation (\ref{r1}) there is no
difference between the Scharzchild and the Kerr black holes.

Now we apply the first law of thermodynamics to a small pitch
$\Delta A$ of the hot stretched membrane \beq d(\rho_{s}\Delta
A)=T_{s}d(\sigma \Delta A)-pd(\Delta A) \eeq Therefore \beq \Delta
A(d\rho_{s}-T_{s}d\sigma )+d(\Delta A)(\rho_{s}-T_{s}\sigma+p)=0
\eeq For the second term \beq d(\Delta A) =d(\Delta \int_{\Delta
A}\sqrt{g_{\phi \phi}g_{\theta \theta}}d\phi d\theta)\simeq{\Delta
(\sqrt{g_{\phi \phi}g_{\theta \theta}}) d\phi d\theta} \eeq and
since $\rho_{s}$ and $\sigma$ is independent of $\phi$ we can get
\beq \rho_{s}=T_{s}\sigma -p, \ d\rho_{s}=T_{s}d\sigma \eeq From
the second relation and (\ref{r1}) we get \beq \label{r2}
\rho_{s}=2cT_{s}^{-1},\ \ \sigma=cT_{s}^{-2} \eeq From the first
relation and (\ref{r2}) we get \beq \label{r3} p=-cT_{s}^{-1} \eeq

Following the Scharzchild black hole case, we can get the same
unusual dispersion relation $E=\frac{m^2}{k}$ which implies the
interpretation that spacetimes on the stretched membrane is
noncommutative.
\section{Noncommutative membrane of
Kerr-Newman Black Holes}

As the Kerr black hole case, the local property of the stretched
membrane of Kerr-Newman Black Hole should be same to the Newman
Black Hole. There are some difficulties to directly obtain the
expected result. But we can still make a conjecture that spacetimes
on the stretched membrane of the Newman Black Hole is
noncommutative.

For Newman Black Hole, the metric is \beq ds^2 = - ( 1 -
\frac{2M}{r} + \frac{Q^2}{r^2} ) dt^2 + ( 1 - \frac{2M}{r} +
\frac{Q^2}{r^2} )^{-1} dr^2 + r^2 d \Omega ^2 \; , \eeq

The local physical quantities of Newman Black Holes are defined as
the Scharzchild Black Hole. After applying the Smarr' mass
formulas for Newman Black Holes and  the first law of
thermodynamics to the stretched membrane, we can easily get \beq
\label{r222} \rho_{s}=2cT_{s}^{-1},\ \ \sigma=cT_{s}^{-2}-2\pi
\Phi_{H} Q (\kappa A)^{-1} \eeq where $\Phi_H$ is the co-rotating
electric potential on the horizon. In the second relation there is
an additional term from which we cannot get the expected
dispersion relation by making analysis of the form of the free
energy density $f$ in (\ref{w}).

But we notice that the free energy density (\ref{w}) is adapted to
the noninteracting relativistic gas. For the stretched membrane of
Newman black holes, the microscopic constituent can interact each
other because of electric charges, and the free energy density $f$
should have a different form. So we can make a conjecture that
 it is a general result for all kinds of black holes
that the spacetimes on the stretched membrane is noncommutative.

\section{Conclusion}
Noncommutative spacetimes should be a natural result of quantum
gravity, especially string theory. In the past years,
Noncommutative spacetimes has found its role in cosmology \cite{7}
, so we believe it can also play an important role in other fields
when quantum gravity effects must be considered.
\section{Acknowledgments}
 This work was supported by a grant of Academia Sinica . We are thankful for the
discussions with Prof. M. Li and Q.G. Huang. We are also grateful
for Prof. R.G. Cai and Fei Wang for valuable suggestions.

\end {document}